\documentstyle[12pt]{article}
\begin{document}


\begin{center}
{\Large{\bf Adiabatic approximation in the second quantized 
formulation}}
\end{center}
\vskip .5 truecm
\centerline{\bf Kazuo Fujikawa}

\vskip .4 truecm
\centerline {\it Institute of Quantum Science, College of 
Science and Technology}
\centerline {\it Nihon University, Chiyoda-ku, Tokyo 101-8308, 
Japan}
\vskip 0.5 truecm

\makeatletter
\@addtoreset{equation}{section}
\def\theequation{\thesection.\arabic{equation}}
\makeatother

\begin{abstract}
Recently there have been some controversies about the criterion 
of the adiabatic approximation. It is shown that an approximate 
diagonalization of the effective Hamiltonian in the second 
quantized formulation gives rise to a reliable and  
unambiguous criterion of the adiabatic approximation. This is 
illustrated for 
the model of Marzlin and Sanders and a model related to the 
geometric phase which can be exactly diagonalized in the 
present sense.
\end{abstract}


\section{Introduction}

Recently there have been some controversies about the criterion 
of the adiabatic approximation~\cite{messiah}, which was 
triggered by an 
interesting model of Marzlin and Sanders~\cite{marzlin}. This 
model gives rise to an apparently nonsensical result on the 
basis of a series of logical steps which appear to be justified 
on the basis of the conventional wisdom of the adiabatic 
approximation.

We here recapitulate the analysis in~\cite{marzlin}.
They start with the evolution operator
\begin{eqnarray}
U(t)=T^{\star}\exp{[-\frac{i}{\hbar}\int^{t}_{0}dt \hat{H}(t)]}
\end{eqnarray}
and define the object
\begin{eqnarray}
\bar{\psi}_{n}(t)=U(t)^{\dagger}v_{n}(0)
\end{eqnarray}
where $\hat{H}(t)v_{n}(t)=E_{n}(t)v_{n}(t)$. 
This object satisfies the exact relation
\begin{eqnarray}
i\hbar\partial_{t}\bar{\psi}_{n}(t)&=&-U(t)^{\dagger}\hat{H}(t)
v_{n}(0)
\nonumber\\
&=&-U(t)^{\dagger}\hat{H}(t)U(t)U(t)^{\dagger}v_{n}(0)\nonumber\\
&=&-U(t)^{\dagger}\hat{H}(t)U(t)\bar{\psi}_{n}(t).
\end{eqnarray}
They then introduce the quantity
\begin{eqnarray}
\phi_{n}(t)=\exp{[\frac{i}{\hbar}\int^{t}_{0}dt E_{n}(t)]}
v_{n}(0)
\end{eqnarray}
which satisfies the relation
\begin{eqnarray}
i\hbar\partial_{t}\phi_{n}(t)&=&-E_{n}(t)\phi_{n}(t)
\nonumber\\
&=&-E_{n}(t)U(t)^{\dagger}U(t)\exp{[\frac{i}{\hbar}\int^{t}_{0}dt E_{n}(t)]}v_{n}(0)\nonumber\\
&\simeq
&-U(t)^{\dagger}\hat{H}(t)U(t)
\exp{[\frac{i}{\hbar}\int^{t}_{0}dt E_{n}(t)]}
v_{n}(0)\nonumber\\
&=&-U(t)^{\dagger}\hat{H}(t)U(t)\phi_{n}(t)
\end{eqnarray}
where we used the conventional adiabatic approximation (diagonal
dominance) for the Hamiltonian $\hat{H}(t)$ in the 
sense~\cite{berry, simon}
\begin{eqnarray}
U(t)v_{n}(0)&=&\sum_{m}v_{m}(t)v^{\dagger}_{m}(t)U(t)v_{n}(0)
\nonumber\\
&\simeq&v_{n}(t)\exp\{-\frac{i}{\hbar}\int^{t}_{0}dt[E_{n}(t)
- v_{n}^{\dagger}(t)i\hbar\partial_{t}v_{n}(t)]\}
\end{eqnarray}
and thus $\hat{H}(t)U(t)v_{n}(0)\simeq E_{n}(t)U(t)v_{n}(0)$,
namely, the state $v_{n}(0)$ which is the eigenstate of 
$\hat{H}(0)$ with the eigenvalue $E_{n}(0)$ at $t=0$ remains the 
eigenstate of $\hat{H}(t)$ with eigenvalue $E_{n}(t)$ for the 
time development defined by $U(t)v_{n}(0)$.  

On the basis of the relation (1.5), one may attempt to 
identify~\cite{marzlin} 
\begin{eqnarray}
\bar{\psi}_{n}(t)_{adiabatic}=\phi_{n}(t).
\end{eqnarray}
The authors of ~\cite{marzlin} then showed on the basis of the 
identification (1.7) and by using (1.6) 
\begin{eqnarray}
v_{n}^{\dagger}(0)v_{n}(0)&=&
v_{n}^{\dagger}(0)U(t)U(t)^{\dagger}v_{n}(0)
\nonumber\\
&=&v_{n}^{\dagger}(0)U(t)\bar{\psi}_{n}(t)\nonumber\\
&\simeq&v_{n}^{\dagger}(0)U(t)\phi_{n}(t)\nonumber\\
&\simeq&
v_{n}^{\dagger}(0)v_{n}(t)\exp{[
i\int_{0}^{t}dt v_{n}^{\dagger}(t)i\partial_{t}v_{n}(t)]}
\nonumber\\
&\neq&1
\end{eqnarray}
which they argued is false. We shall analyze this problem in 
detail in Section 3. 

Since the publication of the paper by Marzlin and 
Sanders~\cite{marzlin}, many papers which attempted to clarify 
the problem appeared\cite{tong, duki, ma, larson,mackenzie,wu,
vertesi,ye, wu2, tong2, pati,sarandy,ambainis, tong3, jansen,
comparat}. Some of these 
papers presented a general and more precise criterion of 
the adiabatic approximation in the first quantization scheme.
But it appears that a more precise criterion means more involved 
conditions and thus the basic simplicity is generally lost.

The purpose of the present paper is to show that a field 
theoretical technique, namely, the second quantization technique
gives a simple, reliable and unambiguous formulation of the 
adiabatic approximation. The adiabatic approximation is defined 
as an approximate diagonalization of the effective Hamiltonian
in the second quantized approach, and thus it is equally 
applicable to the operator formulation and to the path 
integral formulation~\cite{fujikawa,fujikawa2}.
In the second quantized formulation, contrary to the first
quantized formulation, we start with an
exact formulation and apply the adiabatic approximation later
and thus the adiabatic geometric phase appears from an 
approximate diagonalization of the effective Hamiltonian. 
The adiabatic geometric phase is a part of the effective 
Hamiltonian and thus dynamical, and it is topologically 
trivial~\cite{fujikawa}. 
This aspect is quite different from the first quantized
treatment where the adiabatic geometric phase is usually treated
separately from the approximate diagonalization of the 
Hamiltonian.

The second quantized approach has been applied to the 
analyses of  all the geometric phases, namely, the 
adiabatic~\cite{fujikawa}, non-adiabatic~\cite{fujikawa3}, and 
mixed state geometric phases~\cite{fujikawa4}, though no
adiabatic approximation is involved in the last two examples. 
A salient feature is that an exact hidden local symmetry appears
in the Schr\"{o}dinger equation in 
this formulation~\cite{fujikawa2} and the associated holonomy 
controls all the known geometric phases~\cite{fujikawa4}. In 
particular, the non-adiabatic phase is treated without using
the notion of the projective Hilbert space whose consistency
with the superposition principle is not 
obvious~\cite{fujikawa3}.

In the present paper, we first briefly summarize the second 
quantized formulation of 
the adiabatic approximation and then apply the formulation
to the model of Marzlin and Sanders~\cite{marzlin} and to a 
model related to the geometric phase which can be exactly 
diagonalized in the present sense.

\section{Second quantized formulation}

We summarize a second quantized formulation of the adiabatic 
approximation~\cite{fujikawa, fujikawa2}.
We expand the field variable in the second quantization 
\begin{eqnarray}
\hat{\psi}(t)=\sum_{n}\hat{b}_{n}(t)v_{n}(t,\vec{x})
\end{eqnarray}
by using a specific basis set defined by  
\begin{eqnarray}
\hat{H}(t)v_{n}(t,\vec{x})=E_{n}(t)v_{n}(t,\vec{x}).
\end{eqnarray}
When one uses the above expansion in the action
\begin{eqnarray}
S&=&\int_{0}^{T}dt\int d^{3}x[
\hat{\psi}^{\star}(t,\vec{x})i\hbar\frac{\partial}{\partial t}
\hat{\psi}(t,\vec{x})-\hat{\psi}^{\star}(t,\vec{x})
\hat{H}(t)\hat{\psi}(t,\vec{x})]\nonumber\\
&=&\int_{0}^{T}dt\{\sum_{n}
\hat{b}^{\dagger}_{n}(t)i\hbar\partial_{t}\hat{b}_{n}(t)
-\hat{H}_{eff}(t) \}
\end{eqnarray}
one obtains the effective Hamiltonian
\begin{eqnarray}
\hat{H}_{eff}(t)&=&
\sum_{n,m}\hat{b}_{n}^{\dagger}(t)
\int d^{3}x[v^{\dagger}_{n}(t,\vec{x})\hat{H}(t)v_{m}(t,\vec{x})
- v^{\dagger}_{n}(t,\vec{x})i\hbar\partial_{t}v_{m}(t,\vec{x})]
\hat{b}_{m}(t)\nonumber\\
&=&\sum_{n,m}\hat{b}_{n}^{\dagger}(t)[
E_{n}(t)\delta_{nm}
- \int d^{3}x v^{\dagger}_{n}(t,\vec{x})i\hbar\partial_{t}
v_{m}(t,\vec{x})]
\hat{b}_{m}(t)
\end{eqnarray}
with the quantized operators satisfying the equal-time
commutators
$[\hat{b}_{n}(t), \hat{b}^{\dagger}_{m}(t)]_{\mp}
=\delta_{n,m}$, but the Bose or Fermi statistics is not important
 in our application.

The exact Schr\"{o}dinger probability amplitude with
$\psi_{n}(0,\vec{x})=v_{n}(0,\vec{x})$ is defined by (by noting
$i\hbar\partial_{t}\hat{\psi}(t,\vec{x})=\hat{H}(t)
\hat{\psi}(t,\vec{x})$)
\begin{eqnarray}
\psi_{n}(t,\vec{x})&=&\langle0|\hat{\psi}(t,\vec{x})
\hat{b}^{\dagger}_{n}(0)|0\rangle\nonumber\\
&=&\sum_{m} v_{m}(t,\vec{x})
\langle m|T^{\star}\exp\{-\frac{i}{\hbar}\int_{0}^{t}
\hat{{\cal H}}_{eff}(t)dt\}|n\rangle
\end{eqnarray}
which is equal to the amplitude in the first quantization 
\begin{eqnarray}
\psi_{n}(t,\vec{x})
&=&\langle \vec{x}|T^{\star}
\exp\{-\frac{i}{\hbar}\int_{0}^{t}
\hat{H}(t)dt \}|n(0)\rangle\nonumber\\
&=&\sum_{m}v_{m}(t,\vec{x})
\langle m(t)|T^{\star}\exp\{-\frac{i}{\hbar}\int_{0}^{t}
\hat{H}(t)dt \}|n(0)\rangle
\end{eqnarray}
if one notes the equality~\cite{fujikawa, fujikawa2}
\begin{eqnarray}
\langle m|T^{\star}\exp\{-\frac{i}{\hbar}\int_{0}^{t}
\hat{{\cal H}}_{eff}(t)dt\}|n\rangle
=\langle m(t)|T^{\star}\exp\{-\frac{i}{\hbar}\int_{0}^{t}
\hat{H}(t)dt \}|n(0)\rangle
\end{eqnarray}
where the state $|m\rangle$ on the left-hand side is defined by 
$\hat{b}^{\dagger}_{m}(0)|0\rangle$ and the state on the 
right-hand side is defined by 
$\langle\vec{x}|m(t)\rangle=v_{m}(t,\vec{x})$, respectively, and 
the Schr\"{o}dinger picture
$\hat{{\cal H}}_{eff}(t)$ is defined 
by setting all $\hat{b}_{n}(t)\rightarrow \hat{b}_{n}(0)$ in 
$\hat{H}_{eff}(t)$. The symbol $T^{\star}$ stands for the time
ordering.
A salient feature of the second quantization is that the 
general ``geometric terms'' 
$\int d^{3}x v^{\dagger}_{n}(t,\vec{x})i\hbar\partial_{t}
v_{m}(t,\vec{x})$ automatically appear in the {\em exact} 
$\hat{{\cal H}}_{eff}(t)$.

The adiabaticity means that the probability
amplitude starting with $v_{n}(0,\vec{x})$ at $t=0$ stays 
in the state $v_{n}(t,\vec{x})$ for any later 
time~\cite{messiah}.
This is equivalent to the statement that 
$\hat{{\cal H}}_{eff}(t)$ is diagonal for each time $t$.
The adiabatic approximation in the second quantization is 
thus defined by the diagonal dominance in 
$\hat{{\cal H}}_{eff}(t)$ which is ensured if 
any difference of the diagonal elements are much bigger
than off-diagonal elements, namely, 
\begin{eqnarray}
&&|v^{\dagger}_{n^{\prime}}(t)i\hbar\partial_{t}
v_{m^{\prime}}(t)|\nonumber\\
&&\ll|(E_{n}(t)- \int d^{3}x v^{\dagger}_{n}(t)
i\hbar\partial_{t}v_{n}(t))-(E_{m}(t)- \int d^{3}x 
v^{\dagger}_{m}(t)i\hbar\partial_{t}v_{m}(t))|
\end{eqnarray}
and 
\begin{eqnarray}
|v^{\dagger}_{n^{\prime}}(t)i\hbar\partial_{t}
v_{m^{\prime}}(t)|
\ll|E_{n}(t)- \int d^{3}x v^{\dagger}_{n}(t)
i\hbar\partial_{t}v_{n}(t)|
\end{eqnarray}
for any $n\neq m$ and $n^{\prime}\neq m^{\prime}$. In this case 
one can approximately diagonalize the above effective Hamiltonian
\begin{eqnarray}
\hat{{\cal H}}_{eff}(t)&\simeq&\sum_{n}\hat{b}_{n}^{\dagger}(0)[
E_{n}(t)\delta_{nm}- \int d^{3}x v^{\dagger}_{n}(t,\vec{x})
i\hbar\partial_{t}v_{n}(t,\vec{x})]
\hat{b}_{n}(0)
\end{eqnarray}
and thus the Schr\"{o}dinger probability amplitude is 
approximately given by~\cite{berry} 
\begin{eqnarray}
\psi_{n}(t,\vec{x})&\simeq&
v_{n}(t,\vec{x})\exp\{-\frac{i}{\hbar}
\int_{0}^{t}dt[E_{n}(t)-\int d^{3}x v^{\dagger}_{n}(t,\vec{x})
i\hbar\partial_{t}v_{n}(t,\vec{x})]\}
\end{eqnarray}
with $\psi_{n}(0,\vec{x})=v_{n}(0,\vec{x})$. This statement is 
accurate for a 
finite number of degrees of freedom with $n=1\sim N$, and when 
$N\rightarrow\infty$ one needs to estimate carefully an infinite 
sum of small off-diagonal elements.
The condition (2.8) ensures that the transition between 
different eigenstates is small and the condition (2.9) ensures
that the diagonal element represents the total phase 
such as in (2.11) accurately. In most cases, the condition (2.9)
is trivially satisfied since one can adjust the origin of the 
energy eigenvalue at will by adding a constant. This adjustment 
of the energy eigenvalue does not influence the geometric 
phase since all the geometric phases are defined as the 
holonomy of the basis vectors~\cite{fujikawa4}, for example, 
\begin{eqnarray}
\tilde{v}_{n}(0,\vec{x})^{\dagger}\tilde{v}_{n}(T,\vec{x})=
v_{n}(0,\vec{x})^{\dagger}v_{n}(T,\vec{x})\exp\{\frac{i}{\hbar}
\int_{0}^{T}dt \int d^{3}x v^{\dagger}_{n}(t,\vec{x})
i\hbar\partial_{t}v_{n}(t,\vec{x})]\},
\end{eqnarray}
in (2.11) for a cyclic evolution with $v_{n}(0,\vec{x})=v_{n}(T,\vec{x})$. Here we defined 
\begin{eqnarray}
\tilde{v}_{n}(t,\vec{x})=v_{n}(t,\vec{x})\exp\{\frac{i}{\hbar}
\int_{0}^{t}dt \int d^{3}x v^{\dagger}_{n}(t,\vec{x})
i\hbar\partial_{t}v_{n}(t,\vec{x})]\}
\end{eqnarray}
which satisfies the parallel
transport condition $\int d^{3}x\tilde{v}_{n}(t,\vec{x})^{\dagger}\partial_{t}\tilde{v}_{n}(t,\vec{x})=0$.

We emphasize that this diagonal dominance or approximate 
diagonalization of the effective Hamiltonian is a precise 
restatement of 
the {\em conventional} idea of the adiabatic approximation.

In passing, we note an exact local (i.e., time-dependent)  
symmetry~\cite{fujikawa2}
\begin{eqnarray}
&&v_{n}(t,\vec{x})\rightarrow v^{\prime}_{n}(t, \vec{x})=
e^{i\alpha_{n}(t)}v_{n}(t,\vec{x}),\nonumber\\
&&\hat{b}_{n}(t) \rightarrow \hat{b}^{\prime}_{n}(t)=
e^{-i\alpha_{n}(t)}\hat{b}_{n}(t), \ \ \ \ n=1,2,3,...,
\end{eqnarray}
in the operator $\hat{\psi}(t,\vec{x})$ in (2.1), which arises
from an arbitrariness in the choice of the coordinates in the 
functional space. Under this local symmetry the 
Schr\"{o}dinger amplitude $\psi_{n}(t,\vec{x})
=\langle0|\hat{\psi}(t,\vec{x})\hat{b}^{\dagger}_{n}(0)|0\rangle$ is transformed as
$ \psi^{\prime}_{n}(t,\vec{x})=e^{i\alpha_{n}(0)}
\psi_{n}(t,\vec{x})$
for any $t$, which corresponds to the ray representation.
We thus find an enormous exact local symmetry behind the 
ray representation, and 
this local symmetry is responsible for the holonomy appearing 
in all the geometric phases~\cite{fujikawa4}.

\section{Applications of the formulation}

\subsection{Model of Marzlin and Sanders}

We now analyze the model introduced by Marzlin and 
Sanders~\cite{marzlin} which is defined by the Hamiltonian
(see eq.(1.3)) 
\begin{eqnarray}
\hat{\bar{H}}=-U(t)^{\dagger}\hat{H}(t)U(t)
\end{eqnarray}
in the second quantized formulation of the adiabatic 
approximation. We expand the field variable as
\begin{eqnarray}
\hat{\psi}(t)=\sum_{n}\hat{b}_{n}(t)\bar{v}_{n}(t)
\end{eqnarray}
where $\bar{v}_{n}(t)$ is defined by
\begin{eqnarray}
\bar{v}_{n}(t)=U(t)^{\dagger}v_{n}(t).
\end{eqnarray}
The basis vectors $\{v_{n}(t)\}$ are defined for $\hat{H}(t)$
by $\hat{H}(t)v_{n}(t)=E_{n}(t)v_{n}(t)$, and thus the basis set 
$\{\bar{v}_{n}(t)\}$ satisfy
\begin{eqnarray}
\hat{\bar{H}}\bar{v}_{n}(t)=
-U(t)^{\dagger}\hat{H}(t)U(t)\bar{v}_{n}(t)=-E_{n}(t)\bar{v}_{n}(t).
\end{eqnarray}
In this section, we consider the problem where the spatial
coordinates $\vec{x}$ do not appear explicitly~\cite{marzlin}.
We then have the effective Hamiltonian in the second quantized 
formulation
\begin{eqnarray}
\hat{H}_{eff}(t)&=&\sum_{n,m}\hat{b}_{n}^{\dagger}(t)[
\bar{v}^{\dagger}_{n}(t)
\hat{\bar{H}}\bar{v}_{m}(t)
-\bar{v}^{\dagger}_{n}(t)i\hbar\partial_{t}\bar{v}_{m}(t)
]\hat{b}_{m}(t)\nonumber\\
&=&\sum_{n,m}\hat{b}_{n}^{\dagger}(t)[
-E_{n}\delta_{n,m}
-\bar{v}^{\dagger}_{n}(t)i\hbar\partial_{t}\bar{v}_{m}(t)
]\hat{b}_{m}(t).
\end{eqnarray}
The off-diagonal terms in the geometric terms are evaluated as 
\begin{eqnarray}
\bar{v}^{\dagger}_{m}(t)i\partial_{t}\bar{v}_{n}(t)
&=&\bar{v}^{\dagger}_{m}(t)
[-\frac{1}{\hbar}U(t)^{\dagger}\hat{H}(t)v_{n}(t)+
U(t)^{\dagger}i\partial_{t}v_{n}(t)]\nonumber\\
&=&-\frac{1}{\hbar}v^{\dagger}_{m}(t)E_{n}(t)v_{n}(t)
+v^{\dagger}_{m}(t)i\partial_{t}v_{n}(t)\nonumber\\
&=&v^{\dagger}_{m}(t)i\partial_{t}v_{n}(t)
\end{eqnarray}
for $m\neq n$. 
Thus the eigenvalues (up to signature) and the off-diagonal 
terms in the geometric terms agree with those of the original 
system specified by $\hat{H}(t)$, for which we {\em assume} the 
validity of the adiabatic approximation: 
Namely, we assume that not only the {\em naive} criterion of the 
adiabatic approximation for the original system specified by 
$\hat{H}$
\begin{eqnarray}
|v^{\dagger}_{n^{\prime}}(t)i\hbar\partial_{t}v_{m^{\prime}}(t)|
\ll |E_{n}(t)-E_{m}(t)|
\end{eqnarray}
but also the precise conditions (2.8) and (2.9) are always 
satisfied for 
any $n\neq m$ and $n^{\prime}\neq m^{\prime}$. In the present 
problem defined by $\hat{\bar{H}}$,
the naive criterion (3.7) is satisfied and thus one 
might expect that the adiabatic approximation may be valid in 
the present problem  also.

We now examine the diagonal elements of the geometric terms
\begin{eqnarray}
\bar{v}^{\dagger}_{n}(t)i\hbar\partial_{t}\bar{v}_{n}(t)
&=&\bar{v}^{\dagger}_{n}(t)[-U(t)^{\dagger}\hat{H}(t)v_{n}(t)+
U(t)^{\dagger}i\hbar\partial_{t}v_{n}(t)]\nonumber\\
&=&-v^{\dagger}_{n}(t)E_{n}(t)v_{n}(t)
+v^{\dagger}_{n}(t)i\hbar\partial_{t}v_{n}(t)\nonumber\\
&=&-E_{n}(t)+v^{\dagger}_{n}(t)i\hbar\partial_{t}v_{n}(t).
\end{eqnarray}
The above effective Hamiltonian (3.5) is thus re-written as 
\begin{eqnarray}
\hat{H}_{eff}(t)&=&\sum_{n,m}\hat{b}_{n}^{\dagger}(t)[
-E_{n}(t)\delta_{nm}
\nonumber\\
&&+E_{n}(t)\delta_{nm}- v^{\dagger}_{n}(t)
i\hbar\frac{\partial}{\partial t}v_{m}(t)]
\hat{b}_{m}(t)\nonumber\\
&=&\sum_{n,m}\hat{b}_{n}^{\dagger}(t)[-v^{\dagger}_{n}(t)
i\hbar\frac{\partial}{\partial t}v_{m}(t)]
\hat{b}_{m}(t).
\end{eqnarray}
The  system introduced in~\cite{marzlin} is quite peculiar.
This system contains only the ``small'' elements in the 
effective Hamiltonian and thus we have 
no reliable diagonal dominance, i.e., no reliable adiabatic 
approximation for $\hat{\bar{H}}(t)$ in the conventional sense.
 The crucial property of the 
present problem is that the condition (2.8) for the system
specified by $\hat{\bar{H}}(t)$ 
\begin{eqnarray}
&&|\bar{v}^{\dagger}_{n^{\prime}}(t)i\hbar\partial_{t}
\bar{v}_{m^{\prime}}(t)|\nonumber\\
&&\ll |(-E_{n}(t)- \bar{v}^{\dagger}_{n}(t)
i\hbar\partial_{t}\bar{v}_{n}(t))-(-E_{m}(t)
-\bar{v}^{\dagger}_{m}(t)
i\hbar\partial_{t}\bar{v}_{m}(t))|  
\end{eqnarray}
for any $n\neq m$ and $n^{\prime}\neq m^{\prime}$ is not 
satisfied.

If it happens that
\begin{eqnarray}
&&|v^{\dagger}_{n^{\prime}}(t)i\hbar\partial_{t}
v_{m^{\prime}}(t)|\ll |
v^{\dagger}_{n}(t)i\hbar\partial_{t}v_{n}(t)
-v^{\dagger}_{m}(t)i\hbar\partial_{t}v_{m}(t)|,\nonumber\\
&&|v^{\dagger}_{n^{\prime}}(t)i\hbar\partial_{t}
v_{m^{\prime}}(t)|\ll |
v^{\dagger}_{n}(t)i\hbar\partial_{t}v_{n}(t)|
\end{eqnarray}
for any $n\neq m$ and $n^{\prime}\neq m^{\prime}$ in (3.9), 
however,
one can define a reliable adiabatic approximation for the above 
$\hat{H}_{eff}(t)$. The second condition in (3.11) corresponds 
to (2.9). 
We thus examine the possibility (3.11). In this 
case we have from (2.5)
\begin{eqnarray}
\bar{\psi}_{n}(t)&=&\sum_{m}\bar{v}_{m}(t)\langle m|T^{\star}
\exp[-\frac{i}{\hbar}
\int_{0}^{t}dt \hat{{\cal H}}_{eff}(t)]|n\rangle\nonumber\\
&\simeq&\bar{v}_{n}(t)\exp\{\frac{i}{\hbar}
\int_{0}^{t}dt v^{\dagger}_{n}(t)i\hbar\partial_{t}v_{n}(t)\}
\nonumber\\
&=&\sum_{m}v_{m}(0)v_{m}(0)^{\dagger}U(t)^{\dagger}
v_{n}(t)\exp\{\frac{i}{\hbar}
\int_{0}^{t}dt v^{\dagger}_{n}(t)i\hbar\partial_{t}v_{n}(t)\}
\nonumber\\
&\simeq&v_{n}(0)\exp\{\frac{i}{\hbar}\int_{0}^{t}dt
[E_{n}(t)-v^{\dagger}_{n}(t)i\hbar\partial_{t}v_{n}(t)]\}
\nonumber\\
&&\times\exp\{\frac{i}{\hbar}
\int_{0}^{t}dt v^{\dagger}_{n}(t)i\hbar\partial_{t}v_{n}(t)\}
\nonumber\\
&=&v_{n}(0)\exp\{\frac{i}{\hbar}\int_{0}^{t}dt
E_{n}(t)\}
\end{eqnarray}
where we used the adiabatic approximation (diagonal 
dominance) for the original system specified by $\hat{H}(t)$  
\begin{eqnarray}
v^{\dagger}_{m}(0)U(t)^{\dagger}
v_{n}(t)&=&\left(v_{n}(t)^{\dagger}U(t)
v_{m}(0)\right)^{\star}\nonumber\\
&\simeq&\exp\{\frac{i}{\hbar}\int_{0}^{t}dt
[E_{n}(t)-v^{\dagger}_{n}(t)i\hbar\partial_{t}v_{n}(t)]\}
\delta_{n,m}.
\end{eqnarray}
We thus recover the result (1.7) in (3.12) under the 
conditions (3.11).

The conditions (3.11) imply that 
\begin{eqnarray}
\partial_{t}v_{n}(t)=\sum_{m}v_{m}(t)(v^{\dagger}_{m}(t)
\partial_{t}v_{n}(t))\simeq v_{n}(t)(v^{\dagger}_{n}(t)
\partial_{t}v_{n}(t))
\end{eqnarray}
which in turn implies
\begin{eqnarray}
\partial_{t}[v_{n}(t)\exp\{i\int_{0}^{t}dt v^{\dagger}_{n}(t)
i\partial_{t}v_{n}(t)\}]\simeq 0,
\end{eqnarray}
namely
\begin{eqnarray}
v_{n}(0)\simeq v_{n}(t)\exp\{i\int_{0}^{t}dt v^{\dagger}_{n}(t)
i\partial_{t}v_{n}(t)\}
\end{eqnarray}
and thus the relation (1.8) is {\em not} false under the 
conditions (3.11).  Eq.(3.16) implies that the
geometric phase or holonomy is trivial for a periodic system 
$v_{n}(0)=v_{n}(T)$ with a period $T$. 

In the generic case where the conditions (3.11) are not 
satisfied, we have the exact amplitude
\begin{eqnarray}
\bar{\psi}_{n}(t)&=&\sum_{m}\bar{v}_{m}(t)\langle m|T^{\star}
\exp[-\frac{i}{\hbar}
\int_{0}^{t}dt \hat{{\cal H}}_{eff}(t)]|n\rangle\nonumber\\
&=&\sum_{m}\bar{v}_{m}(t)\bar{v}^{\dagger}_{m}(t)
T^{\star}\exp[-\frac{i}{\hbar}
\int_{0}^{t}dt \hat{\bar{H}}(t)]v_{n}(0)\nonumber\\
&=&T^{\star}\exp[\frac{i}{\hbar}
\int_{0}^{t}dt U(t)^{\dagger}\hat{H}(t)U(t)]v_{n}(0)\nonumber\\
&=&U(t)^{\dagger}v_{n}(0)
\end{eqnarray}
but no reliable adiabatic approximation for the dynamics 
specified by $\hat{\bar{H}}$. The last equality in 
(3.17), which is a result of (1.3), is directly confirmed by 
defining $f(t)=T^{\star}\exp[\frac{i}{\hbar}
\int_{0}^{t}dt U(t)^{\dagger}\hat{H}(t)U(t)]$ and then 
$U(t)f^{\prime}(t)=(i/\hbar)\hat{H}(t)U(t)f(t)
=-U^{\prime}(t)f(t)$, 
namely, $U(t)f(t)=constant=1$.

One may attempt to rewrite the amplitude (3.17) as
\begin{eqnarray}
\bar{\psi}_{n}(t)&=&\sum_{k,m}v_{k}(0)
\left(v^{\dagger}_{k}(0)U(t)^{\dagger}
v_{m}(t)\right)v^{\dagger}_{m}(t)v_{n}(0)\nonumber\\
&\simeq&\sum_{m}v_{m}(0)\exp\{\frac{i}{\hbar}\int_{0}^{t}dt
[E_{m}(t)-v^{\dagger}_{m}(t)i\hbar\partial_{t}v_{m}(t)]\}
\nonumber\\
&&\times
v^{\dagger}_{m}(t)v_{n}(0)
\end{eqnarray}
where we used the diagonal dominance (3.13)
for the system specified by $\hat{H}(t)$,
but no further reliable approximation. 
If the system has a period $T$ in the sense 
$\hat{H}(T)=\hat{H}(0)$, however, one has a simpler 
expression after one cycle by using 
$v_{n}(0)=v_{n}(T)$ and $v^{\dagger}_{m}(T)v_{n}(0)
=v^{\dagger}_{m}(T)v_{n}(T)=\delta_{m,n}$ in (3.18)
\begin{eqnarray}
\bar{\psi}_{n}(T)&\simeq&v_{n}(0)
\exp\{\frac{i}{\hbar}\int_{0}^{T}dt
[E_{n}(t)-v^{\dagger}_{n}(t)i\hbar\partial_{t}v_{n}(t)]\}
\nonumber\\
&=&v_{n}(T)\exp\{\frac{i}{\hbar}\int_{0}^{T}dt
[E_{n}(t)-v^{\dagger}_{n}(t)i\hbar\partial_{t}v_{n}(t)]\}
\end{eqnarray}
which has the same form as the conventional adiabatic 
approximation for the system specified by $\hat{H}(t)$, except 
for the reversed signature in the exponential. This result (3.19)
is consistent with (3.12) for $t=T$ if one recalls that (3.12) 
is valid only under the condition (3.16).
It is important that the basis vectors $v_{n}(t)$ are defined 
for $\hat{H}(t)$ and not for $\hat{\bar{H}}(t)$, and thus (3.19)
 is not called an adiabatic approximation for the dynamics 
defined by $\hat{\bar{H}}(t)$.\\
\\
{\bf Second model of Marzlin and Sanders}\\

We here briefly comment on the second ``counter example'' 
in~\cite{marzlin}. They consider a two-level system with  
exact time evolution defined by 
\begin{eqnarray}
U(t)=\exp\{-i\omega_{0}t{\bf n}(t)\cdot{\bf \sigma} \} 
\end{eqnarray}
with 
\begin{eqnarray}
{\bf n}(t)=(\cos(2\pi t/\tau),\sin(2\pi t/\tau),0)
\end{eqnarray}
and ${\bf \sigma}=(\sigma_{x},\sigma_{y},\sigma_{z})$ denoting
the Pauli matrices.
Although it is not clearly stated in~\cite{marzlin}, it is 
natural to understand the above evolution operator standing 
for 
\begin{eqnarray}
U(t_{2},t_{1})=\exp\{-i\omega_{0}(t_{2}-t_{1})
{\bf n}(t_{2}-t_{1})\cdot{\bf \sigma} \}. 
\end{eqnarray}
One can then confirm that the above operator does not 
satisfy the basic composition law of quantum mechanics
\begin{eqnarray}
U(t_{3},t_{1})=U(t_{3},t_{2})U(t_{2},t_{1})
\end{eqnarray}
except for the time independent ${\bf n}(t)$. Since their 
operator $U(t)$ does not depend on the intermediate time, the
condition (3.23) implies 
\begin{eqnarray}
U(2t)=U(t)U(t)
\end{eqnarray}
if one chooses $t_{3}-t_{2}=t_{2}-t_{1}=t$. This relation is 
satisfied only for the time independent ${\bf n}(t)$. The 
second model in~~\cite{marzlin} does not constitute 
a meaningful counter example of the adiabatic approximation. 

One may instead start with their Hamiltonian~\cite{marzlin}  
\begin{eqnarray}
\hat{H}(t)={\bf R}(t)\cdot{\bf \sigma}
\end{eqnarray}
where
\begin{eqnarray}
&&R_{x}(t)\equiv|R(t)|\sin\Theta(t)\cos\varphi(t)=
\omega_{0}\cos(\omega t)-\frac{1}{2}\omega\sin(\omega t)
\sin(2\omega_{0}t),\nonumber\\
&&R_{y}(t)\equiv|R(t)|\sin\Theta(t)\sin\varphi(t)=
\omega_{0}\sin(\omega t)+\frac{1}{2}\omega\cos(\omega t)
\sin(2\omega_{0}t),\nonumber\\
&&R_{z}(t)\equiv|R(t)|\cos\Theta(t)=\omega\sin^{2}(\omega_{0}t)
\end{eqnarray}
with $\omega=2\pi/\tau$ and 
\begin{eqnarray}
|R(t)|=\sqrt{\omega^{2}_{0}+\omega^{2}\sin^{2}(\omega_{0}t)},
\end{eqnarray}
without asking where it came from. (If the composition law 
(3.23) is satisfied, one can define the Hamiltonian by 
$U(t_{2}+\Delta t, t_{2})=\exp\{-i\hat{H}(t_{2})\Delta t\}$. 
Since the composition law is  not satisfied by the present 
example, the Hamiltonian thus defined does not agree with 
(3.25) except for the time independent ${\bf n}(t)$. )

One can then construct the instantaneous eigenvectors
\begin{eqnarray}
v_{+}(R)=\left(\begin{array}{c}
            \cos\frac{\Theta}{2}e^{-i\varphi}\\
            \sin\frac{\Theta}{2}
            \end{array}\right), \ \ \ \ \ 
v_{-}(R)=\left(\begin{array}{c}
            \sin\frac{\Theta}{2}e^{-i\varphi}\\
            -\cos\frac{\Theta}{2}
            \end{array}\right)
\end{eqnarray}
as
\begin{eqnarray} 
\hat{H}(t) v_{\pm}(R)=\pm|R(t)|v_{\pm}(R).
\end{eqnarray}
By denoting $m$ and $n$ to run over $\pm$, we define
$v^{\dagger}_{m}(R)i\frac{\partial}{\partial t}v_{n}(R)
=\langle m|i\frac{\partial}{\partial t}|n\rangle$ and 
then
\begin{eqnarray}
\langle +|i\frac{\partial}{\partial t}|+\rangle
&=&\frac{(1+\cos\Theta)}{2}\dot{\varphi},
\nonumber\\
\langle +|i\frac{\partial}{\partial t}|-\rangle
&=&\frac{\sin\Theta}{2}\dot{\varphi}+\frac{i}{2}\dot{\Theta}
=(\langle -|i\frac{\partial}{\partial t}|+\rangle)^{\star}
,\nonumber\\
\langle -|i\frac{\partial}{\partial t}|-\rangle
&=&\frac{1-\cos\Theta}{2}\dot{\varphi}.
\end{eqnarray}
When one expands $\hat{\psi}(t)$ as
\begin{eqnarray}
\hat{\psi}(t)=\sum_{n}\hat{b}_{n}(t)v_{n}(t),
\end{eqnarray}
the exact second quantized effective Hamiltonian (2.4) is given 
by 
\begin{eqnarray}
\hat{H}_{eff}(t)&=&|R(t)|\hat{b}^{\dagger}_{+}
\hat{b}_{+}
-|R(t)|\hat{b}^{\dagger}_{-}\hat{b}_{-}
 -\hbar \sum_{m,n}\hat{b}^{\dagger}_{m}
\langle m|i\frac{\partial}{\partial t}|n\rangle\hat{b}_{n}.
\end{eqnarray}
One can confirm that the diagonal dominance (or conventional 
adiabatic approximation) perfectly works for 
\begin{eqnarray}
\omega_{0}=2n\omega\gg \omega
\end{eqnarray}
with an integer $n$, which is the assumption made 
in~\cite{marzlin}. One may note $|R(t)|\simeq \omega_{0}$,
$0\leq \cos\Theta(t)\ll 1$, $\dot{\Theta}(t)\sim \omega$ and 
$\dot{\varphi}(t)\sim \omega$ in (3.26).

One can also confirm that the evolution operator defined in 
(2.7)
\begin{eqnarray}
\langle m|T^{\star}\exp\{-\frac{i}{\hbar}\int_{0}^{t}
\hat{{\cal H}}_{eff}(t)dt\}|n\rangle
=\langle m(t)|T^{\star}\exp\{-\frac{i}{\hbar}\int_{0}^{t}
\hat{H}(t)dt \}|n(0)\rangle
\end{eqnarray}
satisfies the basic composition law (3.23).

\subsection{Exactly solvable model}

We next study the model described by  
\begin{eqnarray}
\hat{H}(t)=-\mu\hbar\vec{B}(t)\vec{\sigma}
\end{eqnarray}
where $\vec{\sigma}$ stand for Pauli matrices and  
\begin{eqnarray}
\vec{B}(t)=B(\sin\theta\cos\varphi(t), 
\sin\theta\sin\varphi(t),\cos\theta ).
\end{eqnarray}
Here we assume $\varphi(t)=\omega t$ with {\em constant} 
$\omega$, $B$ 
and $\theta$. This model has been studied by various authors
in the past by using the adiabatic 
approximation~\cite{berry, simon}, but to our knowledge, an 
exact treatment was first given 
in Ref.~\cite{fujikawa3}. We here present the 
essence of the analysis with additional comments from the point of
view of the adiabatic approximation.

We have the effective Hamiltonian in (2.4) 
\begin{eqnarray}
\hat{H}_{eff}(t)&=&[-\mu\hbar B
-\frac{(1+\cos\theta)}{2}\hbar\omega]\hat{b}^{\dagger}_{+}
\hat{b}_{+}
+[\mu\hbar B-\frac{1-\cos\theta}{2}\hbar\omega]
\hat{b}^{\dagger}_{-}\hat{b}_{-}
\nonumber\\
&-&\frac{\sin\theta}{2}\hbar\omega
[\hat{b}^{\dagger}_{+}\hat{b}_{-}+
\hat{b}^{\dagger}_{-}\hat{b}_{+}]
\end{eqnarray}
with 
\begin{eqnarray}
v_{+}(t)=\left(\begin{array}{c}
            \cos\frac{1}{2}\theta e^{-i\varphi(t)}\\
            \sin\frac{1}{2}\theta
            \end{array}\right), \ \ \ \ \ 
v_{-}(t)=\left(\begin{array}{c}
            \sin\frac{1}{2}\theta e^{-i\varphi(t)}\\
            -\cos\frac{1}{2}\theta
            \end{array}\right)
\end{eqnarray}
which satisfy $\hat{H}(t)v_{\pm}(t)=\mp\mu\hbar Bv_{\pm}(t)$
and the relations
\begin{eqnarray}
v^{\dagger}_{+}(t)i\frac{\partial}{\partial t}v_{+}(t)
&=&\frac{(1+\cos\theta)}{2}\omega
\nonumber\\
v^{\dagger}_{+}(t)i\frac{\partial}{\partial t}v_{-}(t)
&=&\frac{\sin\theta}{2}\omega
=v^{\dagger}_{-}(t)i\frac{\partial}{\partial t}v_{+}(t)
,\nonumber\\
v^{\dagger}_{-}(t)i\frac{\partial}{\partial t}v_{-}(t)
&=&\frac{1-\cos\theta}{2}\omega.
\end{eqnarray}
We next perform a unitary transformation
\begin{eqnarray}
\left(\begin{array}{c}
     \hat{b}_{+}(t)\\
     \hat{b}_{-}(t)
     \end{array}\right)
&=&
\left(\begin{array}{cc}
 \cos\frac{1}{2}\alpha&-\sin\frac{1}{2}\alpha\\
 \sin\frac{1}{2}\alpha &\cos\frac{1}{2}\alpha
            \end{array}\right)
\left(\begin{array}{c}
           \hat{c}_{+}(t)\\
           \hat{c}_{-}(t)
            \end{array}\right)\nonumber\\
&\equiv&U^{T}\left(\begin{array}{c}
           \hat{c}_{+}(t)\\
           \hat{c}_{-}(t)
            \end{array}\right)
\end{eqnarray}
where $U^{T}$ stands for the transpose of $U$. 
The eigenfunctions are transformed to
\begin{eqnarray}
\left(\begin{array}{c}
     w_{+}(t)\\
     w_{-}(t)
     \end{array}\right)
&=&
\left(\begin{array}{cc}
 \cos\frac{1}{2}\alpha&\sin\frac{1}{2}\alpha\\
 -\sin\frac{1}{2}\alpha &\cos\frac{1}{2}\alpha
            \end{array}\right)
\left(\begin{array}{c}
           v_{+}(t)\\
           v_{-}(t)
            \end{array}\right)
\end{eqnarray}
or explicitly
\begin{eqnarray}
w_{+}(t)=\left(\begin{array}{c}
            \cos\frac{1}{2}(\theta-\alpha) e^{-i\varphi(t)}\\
            \sin\frac{1}{2}(\theta-\alpha)
            \end{array}\right), \ \ \ \ \ 
w_{-}(t)=\left(\begin{array}{c}
            \sin\frac{1}{2}(\theta-\alpha) e^{-i\varphi(t)}\\
            -\cos\frac{1}{2}(\theta-\alpha)
            \end{array}\right).
\end{eqnarray}
The field variable $\hat{\psi}(t,\vec{x})$ in the second 
quantization is given by 
\begin{eqnarray}
\hat{\psi}(t,\vec{x})&=&\sum_{n=\pm}\hat{b}_{n}(t)v_{n}(t)
=\sum_{n=\pm}\hat{c}_{n}(t)w_{n}(t).
\end{eqnarray}
We also have
\begin{eqnarray}
&&w_{\pm}^{\dagger}(t)\hat{H}w_{\pm}(t)
=\mp \mu\hbar B\cos\alpha\nonumber\\
&&w_{\pm}^{\dagger}(t)i\hbar\partial_{t}w_{\pm}(t)
=\frac{\hbar\omega}{2}(1\pm\cos(\theta-\alpha)).
\end{eqnarray}

If one chooses the constant parameter $\alpha$ in (3.40) as 
\begin{eqnarray}
\tan\alpha=\frac{\hbar\omega\sin\theta}{2\mu\hbar B+\hbar\omega
\cos\theta}
\end{eqnarray}
or equivalently
$2\mu\hbar B\sin\alpha=\hbar\omega\sin(\theta-\alpha)$, 
one obtains a diagonal effective Hamiltonian
\begin{eqnarray}
\hat{H}_{eff}(t)&=&\hat{c}^{\dagger}_{+}(t)
[-\mu\hbar B\cos\alpha
-\frac{\hbar\omega}{2}(1+\cos(\theta-\alpha))]\hat{c}_{+}(t)
\nonumber\\
&+&\hat{c}^{\dagger}_{-}(t)
[+\mu\hbar B\cos\alpha
-\frac{\hbar\omega}{2}(1-\cos(\theta-\alpha))]\hat{c}_{-}(t)
\nonumber\\
&=&\sum_{n=\pm}\hat{c}^{\dagger}_{n}(t)
[w_{n}^{\dagger}(t)\hat{H}w_{n}(t)
-w_{n}^{\dagger}(t)i\hbar\partial_{t}w_{n}(t)]\hat{c}_{n}(t).
\end{eqnarray}
The above unitary transformation is time-independent and thus 
the effective Hamiltonian is not changed 
$\hat{H}_{eff}(b^{\dagger}_{\pm}(t),b_{\pm}(t))
=\hat{H}_{eff}(c^{\dagger}_{\pm}(t),c_{\pm}(t))$.

We thus  have the {\em exact} Schr\"{o}dinger amplitudes in (2.5)
\begin{eqnarray}
\psi_{\pm}(t)
&=&w_{\pm}(t)\exp\{-\frac{i}{\hbar}[\mp\mu\hbar B\cos\alpha
-\frac{\hbar\omega}{2}(1\pm\cos(\theta-\alpha))]t\}
\nonumber\\
&=&w_{\pm}(t)\exp\{-\frac{i}{\hbar}\int_{0}^{t}dt
[w_{\pm}^{\dagger}(t)\hat{H}w_{\pm}(t)
-w_{\pm}^{\dagger}(t)i\hbar\partial_{t}w_{\pm}(t)]\}
\end{eqnarray}
which satisfy the Schr\"{o}dinger equation
\begin{equation}
i\hbar{\partial_{t}}\psi_{\pm}(t)=\hat{H}(t)\psi_{\pm}(t)
\end{equation}
with the Hamiltonian in (3.35). This equation is directly 
confirmed for (3.47).
The amplitudes in (3.47) are periodic with period  
$T=\frac{2\pi}{\omega}$ up to a phase, and they are exact 
and thus valid in a non-adiabatic sense also. From the view 
point of 
the diagonalization of the Hamiltonian, we have not completely
diagonalized the starting Hamiltonian (3.35) since 
$w_{\pm}(t)$ carry certain time-dependence.

The separation of the ``dynamical
phase'' (the first term in the exponential) and the 
geometric phase (the second term in the exponential) in (3.47),
both of which arise from the effective Hamiltonian,  
is achieved by varying the parameters in the 
Hamiltonian, namely, $B$ and $\omega$ in the present case.
The formula (3.47) however shows that both of the ``dynamical
phase'' and the geometric phase depend on these parameters
in a non-trivial way. 

We examine two extreme limits:\\
(i)For the {\em adiabatic} limit $\hbar\omega/(\hbar\mu B)\ll 1$,
which ensures the diagonal dominance in (3.37), we have from 
(3.45)
\begin{eqnarray}
\alpha\simeq[\hbar\omega/2\hbar\mu B]\sin\theta.
\end{eqnarray}
If one sets 
$\alpha=0$ approximately in the exact solution of the 
Schr\"{o}dinger equation (3.47), one recovers the ordinary 
Berry phase~\cite{berry}
\begin{eqnarray}
\psi_{\pm}(T)&\simeq&\exp\{i\pi(1\pm\cos\theta) \}
\exp\{\pm\frac{i}{\hbar}\int_{0}^{T}dt
\mu\hbar B\}v_{\pm}(T)
\end{eqnarray}
with $v_{\pm}$ defined in (3.38). The phase factor 
$\exp\{i\pi(1\pm\cos\theta) \}$ is known to be similar to the 
phase induced by a magnetic monopole located at the origin of 
the parameter space.\\
(ii)For the other limit, namely, {\em non-adiabatic} limit 
$\hbar\mu B/(\hbar\omega)\ll 1$, we have from (3.45)
\begin{eqnarray}
\theta-\alpha\simeq[2\hbar\mu B/\hbar\omega]\sin\theta
\end{eqnarray}
and if one sets $\alpha=\theta$ approximately in the exact
solution (3.47), one obtains the trivial geometric phase
\begin{eqnarray}
\psi_{\pm}(T)
&\simeq&w_{\pm}(T)\exp\{\pm\frac{i}{\hbar}\int_{0}^{T}dt
[\mu\hbar B\cos\theta]\}
\end{eqnarray}
with
\begin{eqnarray}
w_{+}(t)&=&\left(\begin{array}{c}
            e^{-i\varphi(t)}\\
            0
            \end{array}\right), \ \ \   
w_{-}(t)=\left(\begin{array}{c}
            0\\
            -1
            \end{array}\right).
\end{eqnarray}
This shows that the monopole-like singularity is smoothly 
connected to the trivial phase inside the exact solution of the 
Schr\"{o}dinger equation, and thus the geometric phase is 
{\em topologically trivial}~\cite{fujikawa}.

This example shows that the second quantized formulation is 
useful not only in formulating a reliable adiabatic 
approximation but also in allowing an exact treatment in some 
cases.

\section{Conclusion}

It has been shown that the adiabatic approximation in the 
second quantized formulation~\cite{fujikawa, fujikawa2} in the 
sense of an
approximate diagonalization of the effective Hamiltonian provides
a reliable criterion of the adiabatic approximation.
The validity of the approximate diagonal dominance depends on 
the conditions (2.8) and (2.9). The model of Marzlin and 
Sanders~\cite{marzlin} 
reminded us of the importance of the crucial conditions (2.8)
and (2.9); the appearance of the combination $E_{n}(t)-\int d^{3}x v^{\dagger}_{n}(t,\vec{x})i\hbar\partial_{t}
v_{n}(t,\vec{x})$ in the conditions 
implies that the geometric phase\\ 
$\int d^{3}x v^{\dagger}_{n}(t,\vec{x})i\hbar\partial_{t}
v_{n}(t,\vec{x})$ is in fact a part of the energy eigenvalue.
\\

I thank D.M. Tong for calling the work by Marzlin and Sanders
and related works to my attention.

\end{document}